\documentclass[manuscript]{aastex}

\shorttitle{Evolutionary of ``black widow'' pulsars}

\shortauthors{Benvenuto, De Vito \& Horvath}

\begin{document}

\title{Evolutionary trajectories of ultra-compact ``black widow'' pulsars\\
    with very low mass companions}

\author{O. G. Benvenuto, M.A. De Vito }
\affil{Facultad de Ciencias Astron\'omicas y Geof\'\i sicas, Universidad Nacional de La Plata\\
and Instituto de Astrof\'\i sica de La Plata (IALP), CCT-CONICET-UNLP. Paseo del Bosque S/N (B1900FWA), La Plata, Argentina}

\email{obenvenu,adevito@fcaglp.unlp.edu.ar}

\and

\author{J.E. Horvath}
\affil{Instituto de Astronomia, Geof\'\i sica e Ci\^encias Atmosf\'ericas, Universidade de S\~ao Paulo\\
R. do Mat\~ao 1226 (05508-090), Cidade Universit\'aria, S\~ao Paulo SP, Brazil}

\email{foton@astro.iag.usp.br}

\begin{abstract} The existence of millisecond pulsars with planet-mass companions \citep{2011Sci...333.1717B}
in close orbits is challenging from the stellar evolution point of view. We calculate
in detail the evolution of binary systems self-consistently, including mass transfer,
evaporation and irradiation of the donor by X-rays feedback, demonstrating the existence
of a new evolutionary path leading to short periods and compact
donors as required by the observations of PSR~J1719-1438. We also point out the
alternative of an exotic nature of the companion planet-mass star. \end{abstract}


\section{INTRODUCTION}

The recent report \citep{2011Sci...333.1717B} of a binary millisecond pulsar in a
2.2~h orbit featuring a Jupiter-like mass companion with a lower
bound for the mean density of ${\bar \rho} \geq 23\; g\; cm^{-3}$ is
both important and challenging for stellar evolution theory.
Indeed, the role of the pulsar wind and illumination feedback have
been deemed as important \citep{2011Sci...333.1717B}, but it was not clear whether
the interplay of all the effects is enough to reproduce the
observed features. In order to account for the existence of the
PSR~J1719-1438 system we have looked for an evolutionary scenario
in which a normal star evolves losing most of its mass and
reaching the observed configuration. We have considered close
binary systems composed by an accreting neutron star (NS) orbiting together
with a normal donor star. We attempt to answer the full history of these systems below, and
report the first complete results in this work.

\section{CALCULATIONS AND RESULTS}

In the case of this kind of systems, the donor star evolves
essentially as an isolated object up to the moment at which its
radius $R_{2}$  nearly equals the radius of the Roche Lobe
$R_{L}$\footnote{As usual, we shall refer the NS (donor star) as
the primary (secondary) with subindex 1~(2).}. This phenomenon is
usually referred to as the onset of the Roche Lobe Overflow
(RLOF). Near RLOF, tidal dissipation forces the orbit to become
circular with an period $P_{i}$, a starting point for the
calculations. For the case of low mass donor stars with masses
fulfilling the condition $0<M_{2}/M_{1}<0.8$, the radius of a
sphere with the volume of the Roche Lobe can be approximated by
\citep{1971ARA&A...9..183P}

\begin{equation}
R_{L} = 0.46224\; a\; {\biggl( {M_{2}
\over{M_{1}+M_{2}}}\biggr)}^{1/3},
\end{equation}

where $a$ is the semiaxis of the circular orbit. Since that moment
on, the donor star transfers mass across the Lagrangian point
$L_{1}$ towards the NS. This process, in turn, makes the orbit to
evolve. At present it is not clear how much of the matter
transferred from the donor star is effectively accreted by the NS.
Hereafter we define $\beta$ as the fraction of transferred
material that is accreted by the NS (${\dot M}_{1} = -\beta {\dot
M}_{2}$, but always below the Eddington limit ${\dot M}_{Edd} = 2
\times 10^{-8}\; M_{\odot}\; yr^{-1}$; see, e.g.,
\citealt{2002ApJ...565.1107P}). If $\beta < 1$ some material is
lost by the system carrying away the specific angular momentum of
the secondary. As the value of $\beta$ is not critical in
determining the evolution of this kind of systems
\citep{2012MNRAS.421.2206D}, hereafter we shall assume an average
value of $\beta = 1/2$. This value of  $\beta$ has been usually
assumed, for example in \citet{2002ApJ...565.1107P}, and more
recently in population synthesis calculations by
\citet{2008ApJS..174..223B}. Also, gravitational radiation
\citep{1975ctf..book.....L} and magnetic braking
\citep{1981A&A...100L...7V} are known to provide relevant angular
momentum sinks\footnote{Another law of magnetic braking has been
presented in \citet{2003ApJ...599..516I}; adopting it the results
of the present work may change. This will be explored in a future
paper.}.

${\dot M}_{2}$ due to RLOF is described by the expression given by
\citet{1988A&A...202...93R},

\begin{equation}
{\dot M}_{2, RLOF} = -{\dot M}_{0} \exp{\biggl( {R_{2} - R_{L}
\over{H_{P}}} \biggr)} ,
\end{equation}

where ${\dot M}_{0}$ is a smooth function of $M_{1}$ and $M_{2}$,
and $H_{P}$ is the pressure scale height at the photosphere (for
further details, see \citealt{1988A&A...202...93R}). The above
given description corresponds to the standard treatment for the
evolution of Low Mass X-Ray Binaries (LMXBs); see, e.g.,
\citet{2002ApJ...565.1107P}. However, as it will be clear below,
these ingredients are {\it not} enough to account for the
formation of a binary pair like PSR~J1719-1438. Another effect
that drives further mass loss from the donor star is the
evaporating wind, driven by the pulsar radiation. Following
\citet{1992MNRAS.254P..19S}, we include this effect by considering

\begin{equation}
{\dot M}_{2,evap} = - {f \over{2 v_{2,esc}^{2}}} L_{P} {\biggl(
{R_{2} \over{a}} \biggr)}^{2} ,
\end{equation}

where the pulsar's spin down luminosity $L_{P}$ is given by
$L_{P}= 4 \pi^{2} I_{1} P_{1} {\dot P}_{1}$ ($I_{1}$ is the moment
of inertia of the NS, $P_{1}$ is its spin period and ${\dot
P}_{1}$ its spindown rate), $v_{2,esc}$ is the escape velocity
from the donor star surface and $f$ is an efficiency factor that
will be set to 0.1 as in \citet{1992MNRAS.254P..19S}. As we shall
be concerned with very short orbital periods, a relevant
phenomenon to be considered is irradiation feedback. When the
donor star transfers mass onto the NS, it releases an accretion
luminosity that illuminates the donor star with a flux $F_{irr} =
{{\alpha_{irr}} \over{4 \pi a^{2}}} {G M_{1} \over{R_{1}}} {\dot
M_{1}}$, where $\alpha_{irr}$ is a constant that accounts for the
fact that neither all the luminosity has to be released as
electromagnetic radiation, nor has to be emitted isotropically
\citep{2004A&A...423..281B}. The radiation incident onto the donor
star partially blocks the release of its internal energy,
modifying its evolution. This problem has been addressed by
\citet{1997A&AS..123..273H}. Here we shall assume the validity of
the point-source model, i.e., that the accreting NS is the only
source of radiation incident onto the donor star.

In order to compute the evolution of these systems we have
employed our detailed (Henyey) evolutionary code described in
\citet{2003MNRAS.342...50B} and \citet{2012MNRAS.421.2206D}, which
has been modified to incorporate irradiation feedback and
evaporating winds as described above. For a compact binary system
to be an adequate candidate to account for the properties of
PSR~J1719-1438, it must have a very close orbit. The RLOF will
occur during the hydrogen core burning stage; this is usually
classified as a Class A mass transfer episode
\citep{1967ZA.....65..251K} (whereas Class B and C episodes
correspond to the cases in which the onset of the RLOF occurs
after the exhaustion of core hydrogen and helium respectively). An
exploration of the parameter space defining a particular compact
binary system ($M_{1},M_{2},P_{i}$) indicates that there exists a
restricted region of which that leads to the formation of systems
like PSR~J1719-1438. For example, there is a narrow range for the
initial orbital period: if $P_{i}$ is too short (say, $<$ 0.5 d)
even at the minimum radius (on the ZAMS), the donor star would be
transferring mass. On the other side, if $P_{i}$ is larger than
about $\approx$ 0.9 d, the system evolves on an orbit that widens
enough to allow for the formation of a low-mass ($\sim 0.25
M_{\odot}$) helium white dwarf (HeWD) - millisecond pulsar pair
(see, e.g., \citealt{2012MNRAS.421.2206D}). Thus, the values of
$P_{i}$ leading to objects on converging orbits is very
restricted. If the formation of the PSR~J1719-1438 system
proceeded the way we considered here, $P_{i}$ should have fallen
in this interval\footnote{In any case, here we should remark that
the interval of $P_{i}$ referred above depends on the particular
physical ingredients assumed in our computations that are
certainly not fully known. The related present uncertainties
should affect the precise value of the period interval, although
it will be still within a narrow range.}. Furthermore, as we shall
see below, the system has to evolve for a quite a long time to
reach mass values as low as those indicated by observations. This,
in turn, imposes a lower limit for its initial mass: if it is very
low, the system would need to evolve for a time in excess of the
age of the Universe. For more massive stars, there exists a limit
imposed by the stability of the mass transfer at the onset of the
RLOF (for further details see \citealt{2002ApJ...565.1107P}). This
set of conditions strongly suggests that these systems should be
rare.

Observations of the pulsar signal are quite stable, and thus do
not suggest that PSR~J1719-1438 is undergoing a RLOF episode
(\citealt{2011Sci...333.1717B}; M. Bailes, private communication).
Thus, the donor star should be smaller than its corresponding
Roche Lobe. If RLOF were the only process giving rise to mass loss
from the donor star, this fact would be very difficult to account
for. It is well known that low mass WDs behave like polytropic
spheres of index $n = 1.5$. For these structures, the mass-radius
relation is $R \propto M^{-1/3}$. Therefore, the star expands in
response to mass loss. This behavior continues as long as the
equation of state is dominated by electron degeneracy. As the star
experiences further mass loss, it lowers its density and the
degree of degeneracy; and non-ideal effects become more important.
Eventually, the mass-radius relation changes, and there is a mass
value for which the radius passes through a maximum. For example,
if the object has a helium dominated composition this corresponds
to an object with a mass of $M_{2} = 2 \times 10^{-3} M_{\odot}$,
which has a radius of $R_{2} = 5 \times 10^{-2} R_{\odot}$ (see,
e.g., \citealt{2003ApJ...598.1217D}). As the less massive object
in the system losses mass (and angular momentum), we arrive to a
situation in which the orbit gets wider while keeping  $R_{2} -
R_{L} \approx H_{P}$. When the donor star reaches the
maximum-radius mass value, a further mass loss will force the star
to contract,  {\it detaching} the donor star from its Roche lobe
(notice that for these mass values of the donor star, the
timescale of orbital evolution due to gravitational radiation is
too long to lead the donor star into contact).

If mass loss/transfer were only due to RLOF episode(s), in
reaching the observed configuration, the system would need a
timescale in excess of the age of the Universe. A natural way out
of this apparent paradox is provided by the evaporating wind
described above. As a matter of fact, during the advanced stages
of evolution in which the donor star mass becomes very low ($M_{2}
\leq 2 \times 10^{-2} M_{\odot}$), such evaporating wind dominates
the donor star mass losses and the orbital evolution, even if the
system is still on RLOF conditions. This effect makes the orbit to
become wider than it would be if we consider RLOF solely, making
the star to {\it detach} from its Roche lobe before reaching mass
values as low as that corresponding to the maximum radius for its
composition. In order to explore the plausibility of this scenario
we have computed the evolution of several systems assuming a solar
composition donor star with an initial mass value of $M_{2} = 2
M_{\odot}$, a ``canonical'' NS of  $M_{1} = 1.4 M_{\odot}$, and
some values for the initial orbital period that lead to this kind
of binary systems $P_{i} =$ 0.75~d, 0.80~d and 0.85~d. We
considered evolutionary sequences with and without irradiation
feedback. In this paper we shall not discuss the process of
formation of main sequence-NS CBSs from which we star out our
calculations. Also, we should warn the reader that it is expected
to be possible to arrive to a configuration like that of
PSR~J1719-1438 from other initial conditions, different from the
ones we assumed. These processes have been discussed by
\citet{2004ApJ...603..690B} and references therein. The work by
\citet{2012A&A...541A..22V} noticed several problems in the
formation of the system and attempted to model the outcome varying
the donor and its wind.

\begin{figure}
\epsscale{.60} \plotone{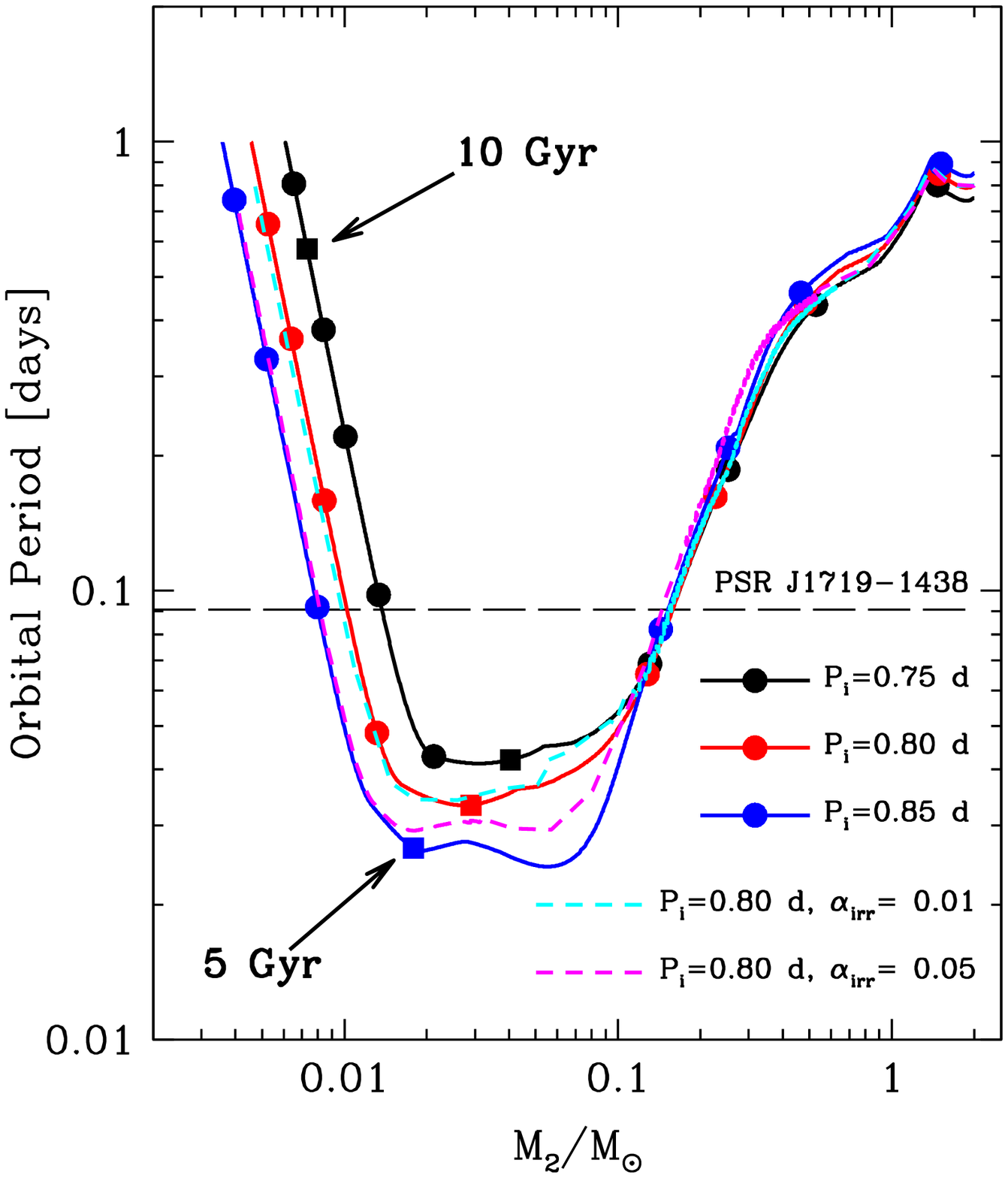}{} \caption{The orbital
period-mass relation for the donor star corresponding to systems
with a solar composition, $2 M_{\odot}$ normal star and a $1.4
M_{\odot}$ neutron star in orbits with initial periods $P_{i}$ of
0.75~d (black), 0.80~d (red), and 0.85~d (blue) respectively. Full
lines correspond to calculations neglecting irradiation feedback,
while the results for 0.80~d and two values for the irradiation
parameter $\alpha_{irr}$  (0.01 -cyan- and 0.05 -pink-) are shown
with dashed lines. Stars spend 1 Gyr evolving leftwards from one
mark to the next one along the trajectories. The observed orbital
period for PSR~J1719-1438 is marked with an horizontal dashed
line. These systems attain the observed period with adequate
masses ($< 0.05 M_{\odot}$) after long (6-7~Gyr), but acceptable
timescales. \label{Fig:periodos}}
\end{figure}

In Fig.~\ref{Fig:periodos} we show the orbital period of the
system as a function of the donor mass. It is remarkable that
irradiation feedback does not induce any dramatic effect on such a
relation. The observed period for PSR~J1719-1438 indicates that
for each model two solutions exist: one with $M_{2} \geq 0.10
M_{\odot}$, (while the orbit is shrinking) and other with $M_{2} =
0.01 M_{\odot}$ (while the orbit is expanding). In view of the
mass function of this system \citep{2011Sci...333.1717B},

\begin{equation}
f(m_{c}) = {4 \pi^{2} \over{G}} {\biggr( {a_{2} \sin i \over{P}}
\biggl)}^{2} = {(M_{2} \sin i)^{3} \over {(M_{1}+M_{2})^{2}}} =
7.85(1) \times 10^{-10} M_{\odot} ,
\end{equation}

\begin{figure}
\epsscale{.60} \plotone{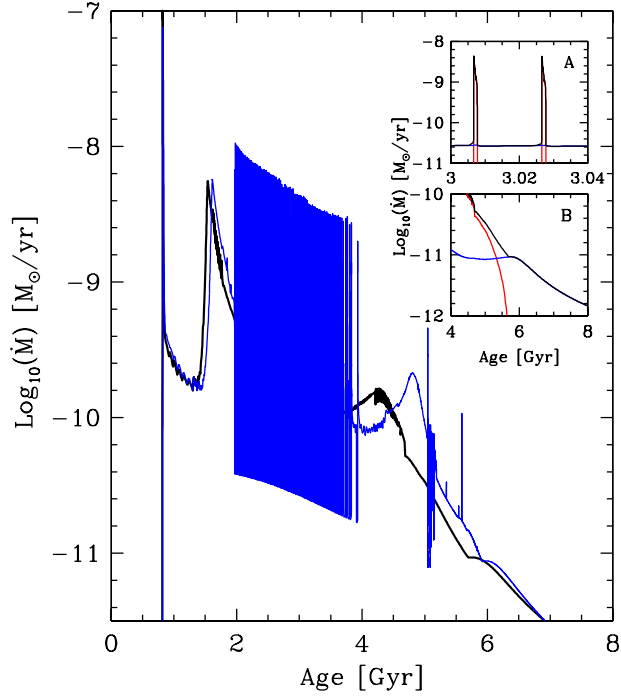}{} \caption{The evolution of
the mass transfer rate from the donor star for the case $P_{i} =$
0.8~d, ignoring (black line) and considering (blue line)
irradiation feedback (with $\alpha_{irr} = 0.05$) as in Fig.~\ref{Fig:periodos}.
The mean value of the mass transfer rate is very similar,
irrespective of the inclusion of irradiation feedback. However,
for ages between 2 and 4~Gyr, irradiated models undergo a sequence
of RLOF episodes similar to those found by \citet{2004A&A...423..281B}.
Some of these episodes are depicted in further detail in
the inset A where black, red and blue lines represent the total
mass transfer rate, the RLOF and evaporation contributions
respectively. Finally, in inset B we show the same quantities
(with colors having the same meaning as in inset A) at the end of
the non-irradiated sequence (irradiated ones behave in a similar
way). Remarkably, after 5.5~Gyr mass transfer rate is dominated by
the evaporation wind driven by pulsar irradiation. \label{Fig:mdots} }
\end{figure}

\begin{figure}
\epsscale{.60} \plotone{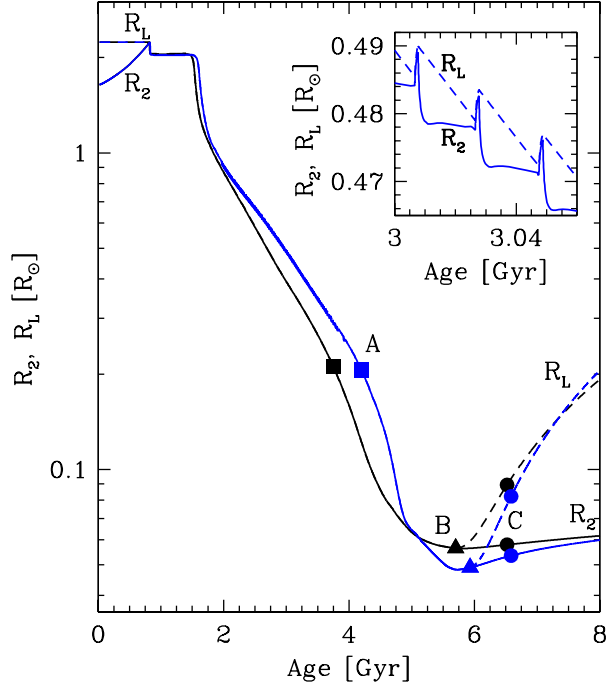}{} \caption{The evolution of
the radius of the donor star $R_{2}$ (solid lines) and its
corresponding Roche lobe $R_{L}$ (dashed lines) for the same
models of  Fig.~\ref{Fig:mdots}, ignoring (black lines) and
considering (blue lines) irradiation feedback (with $\alpha_{irr}
= 0.05$). In the case of the irradiated sequence, the donor star
suffers from a series of contractions and expansions corresponding
to the cyclic mass transfer regime, shown in detail in the inset.
However, the long term evolution of the radii of irradiated models
is very similar to those of non-irradiated ones. Points labelled
with A and C correspond to the stages at which the orbital period
is equal to the observed value, while B indicates the final
detachment of the donor star. Notice that, remarkably, due to the
evaporation wind driven by pulsar irradiation, the donor star
attains the orbital period observed for PSR~J1719-1438 with low
mass values ($M_{2} \approx 0.010 M_{\odot}$) in detached
conditions, as indicated by observations. \label{Fig:radios} }
\end{figure}

\begin{figure}
\epsscale{.60} \plotone{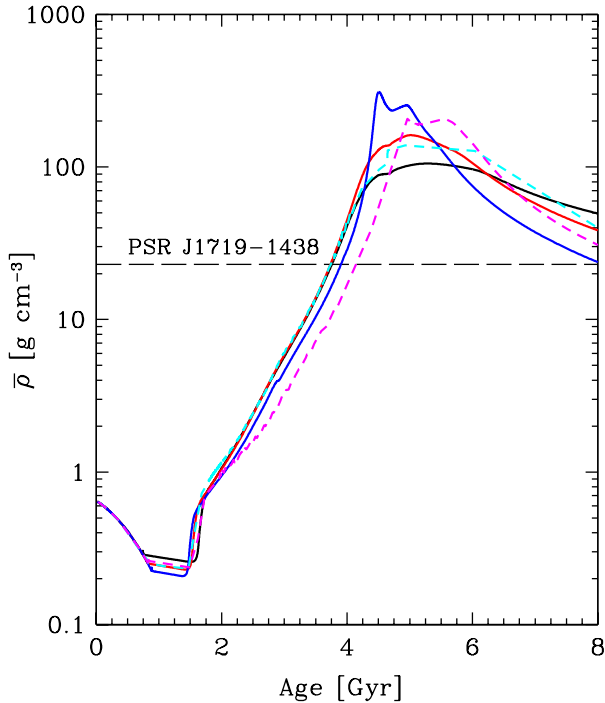}{} \caption{The evolution
of the mean density for the donor star for the same sequences
shown in Fig.~\ref{Fig:periodos}. Lines and colors have the same meaning as there.
We show with a horizontal line the minimum mean density found in
\citet{2011Sci...333.1717B} inferred from observations. First the systems
attain the observed orbital period when under RLOF conditions,
when the mean density is lower than the value ${\bar \rho}= 23\; g\; cm^{-3}$.
After that the star detaches from the Roche lobe and
the mean density increases above the referred minimum value, in
agreement with observations.\label{Fig:rho_prom}}
\end{figure}

(where $P$ is the orbital period, $a_{2}$ is the semiaxis of the
pulsar orbit and $i$ is the inclination angle of the orbit with
respect to the line of sight), it is difficult to consider the
first solution as physically plausible. For $M_{1} = 1.40
M_{\odot}$, we would need $\sin i \approx 0.01$, which has a very
low probability, whereas for the other solution we still need
small but tolerable values of $\sin i \approx 0.1$. Notice that
because at late times the evaporating wind dominates mass
loss/transfer (see below), it accelerates the evolution of the
system (as compared with the standard case in which this effect is
ignored) making it possible to reach a configuration compatible
with the observed state of PSR~J1719-1438 within a long but
acceptable timescale of 6-7~Gyr. We show in Fig.~\ref{Fig:mdots}
the mass transfer rate for the case of $P_{i}=$~0.8~d, with and
without irradiation feedback. Due to irradiation, the donor star
undergoes cyclic mass transfer episodes in a way similar to that
found by \citet{2004A&A...423..281B}. Notice that this oscillating
behavior is restricted to an intermediate stage of evolution.
Despite the uncertainties associated with the present treatment of
irradiation feedback \citep{2008NewAR..51..869R}, it is a
fortunate situation that the final properties of the system are
largely independent of the former. We show in
Fig.~\ref{Fig:radios} the evolution of the donor and the Roche
lobe radii, demonstrating that the system ultimately detaches
around 6~Gyr. Finally, in Fig.~\ref{Fig:rho_prom} we show the
evolution of the mean density of the donor star ${\bar \rho}$. We
find that since ages~$\approx$4~Gyr on (well before detachment
from the Roche lobe),  ${\bar \rho}$ overcomes the lower limit
deduced from observations.

Regarding the final internal composition of the donor star, it
depends on the value of $P_{i}$. For the shortest possible initial
orbital periods, core hydrogen burning is quenched by mass loss
(internal temperature falls down fast enough to appreciably slow
down nuclear activity) and the final hydrogen abundance is
$\approx 0.45$ by mass. For the largest $P_{i}$ for which CBSs
evolve to a black widow configurations, compatible with the
characteristics of the PSR~J1719-1438 system, hydrogen is almost
absent being a helium-dominated composition. It is worth to remark
that for the formation path we addressed in this paper, no other
composition is possible for the donor star interior.

\section{DISCUSSION AND CONCLUSIONS}

The conclusion of this study is to identify a definite new path
for the evolution of binary systems evolving into planet-like -
millisecond pulsars pairs, featuring $R_{2} < R_{L}$ for the donor
star and a mean density ${\bar \rho} > 23 \; g \; cm^{-3}$ for it.
Our calculations show self-consistently that this is indeed
possible, even for objects composed by a mixture of hydrogen and
helium, without the need of postulating a carbon interior, on a
reasonable timescale. The initial conditions for this evolution
are actually quite stringent, as identified above; otherwise the
outcome of the evolution is very different.

Finally, the exciting possibility that the companion of the
millisecond pulsar PSR~J1719-1438 is {\it not} a WD-like but a
truly exotic object (i.e. composed of some form of quark matter)
should not be overlooked. This would easily explain why there is
no modulation even for an edge-on inclination. Actually, in the
latter case the stringent photometric limits derived in
\citet{2011Sci...333.1717B} using Keck-LRIS instrument cannot be
used to place constraints to the inclination, because the absence
of signal is a quite natural outcome. In other words, for the
cases of a strange quark matter nugget or structured strangelet
chunk, the size of the companion would be too small to detect any
photometric signal. The exotic model also predicts that no
carbon/helium lines should be ever observed associated to the
companion. In addition, the lack of detection of evaporation
signatures  \citep{2011Sci...333.1717B} would be naturally
accommodated. Finally, and because of angular momentum
considerations, we expect that the orbit angular momentum
$\vec{J}$ of the quark companion to be aligned with the spin of
the pulsar ``born in original spin'', e.g.,
\citet{1994ApJ...421L..15C}. These are quite strong, albeit
straightforward predictions to be checked in futures studies
addressing the nature of this system. The proposals of extended
exotic stars (strangelet dwarfs, \citealt{2012JPhG...39f5201A};
and strange dwarfs, \citealt{1995ApJ...450..253G}), would need an
evaluation of their surface properties, which would still depend
on the existence or absence of a normal matter atmosphere to
reprocess the incident pulsar radiation. This would be difficult
to distinguish from conventional helium or carbon WDs. However, in
these scenarios there is no link between the evolution of the
system and the final masses and period, and the millisecond pulsar
could be very young and not recycled at all.

\acknowledgments We would like to thank to our referee, Chris
Belczynski, for his prompt and constructive report that helped us
to improve the original version of this paper. O.G.B is member of
the Carrera de Investigador of the CIC-PBA Agency and M.A.D.V. is
member of the Carrera de Investigador, CONICET, Argentina. J.E.H.
has been supported by Fapesp (S\~ao Paulo, Brazil) and CNPq,
Brazil funding agencies.


\clearpage


\clearpage

\clearpage

\clearpage

\clearpage

\begin{deluxetable}{cc}
\tabletypesize{\scriptsize}
\tablecaption{Parameters of the PSR~J1719-1438 employed in the calculations
(from \citet{2011Sci...333.1717B}}
\tablewidth{0pt}
\tablehead{\colhead{Parameter} & \colhead{Value} }
\startdata
$\nu (s^{-1})$        & 172.70704459860(3)~Hz    \\
${\dot \nu} (s^{-2})$ & $-2.2(2)\times 10^{-16}$ \\
Epoch (MJD)           & 55411.0   \\
$P_{orb}$ (d)         & 0.090706293(2) \\
$a_{p} \sin i$ (lt-s) & 0.001819(1)\\
${\bar \rho} (g\; cm^{-3})$ (inferred)& $\geq 23$\\
\enddata

\end{deluxetable}


\begin{thebibliography}{}

\bibitem[Alford et al.(2012)]{2012JPhG...39f5201A} Alford, M.~G., Han, S.,
\& Reddy, S.\ 2012, Journal of Physics G Nuclear Physics, 39, 065201

\bibitem[Bailes et al.(2011)]{2011Sci...333.1717B} Bailes, M., Bates,  S.~D.,
Bhalerao, V., et al.\ 2011, Science, 333, 1717

\bibitem[Belczynski \& Taam(2004)]{2004ApJ...603..690B} Belczynski, K., \& Taam, R.~E.\ 2004, \apj, 603, 690

\bibitem[Belczynski et al.(2008)]{2008ApJS..174..223B} Belczynski, K.,
Kalogera, V., Rasio, F.~A., et al.\ 2008, \apjs, 174, 223

\bibitem[Benvenuto  \& De Vito(2003)]{2003MNRAS.342...50B} Benvenuto, O.~G., \&
De Vito, M.~A.\ 2003, \mnras, 342, 50

\bibitem[B{\"u}ning  \& Ritter(2004)]{2004A&A...423..281B} B{\"u}ning, A., \&
Ritter, H.\ 2004, \aap, 423, 281

\bibitem[Camilo et al.(1994)]{1994ApJ...421L..15C} Camilo, F., Thorsett,  S.~E.,
\& Kulkarni, S.~R.\ 1994, \apjl, 421, L15

\bibitem[Deloye  \& Bildsten(2003)]{2003ApJ...598.1217D} Deloye, C.~J., \&
Bildsten, L.\ 2003, \apj, 598, 1217

\bibitem[De Vito  \& Benvenuto(2012)]{2012MNRAS.421.2206D} De Vito, M.~A., \&
Benvenuto, O.~G.\ 2012, \mnras, 421, 2206

\bibitem[Glendenning et al.(1995)]{1995ApJ...450..253G} Glendenning, N.~K.,
Kettner, C., \& Weber, F.\ 1995, \apj, 450, 253

\bibitem[Hameury  \& Ritter(1997)]{1997A&AS..123..273H} Hameury, J.-M., \&
Ritter, H.\ 1997, \aaps, 123, 273

\bibitem[Ivanova \& Taam(2003)]{2003ApJ...599..516I} Ivanova, N., \& Taam, R.~E.\ 2003, \apj, 599, 516

\bibitem[Kippenhahn  \& Weigert(1967)]{1967ZA.....65..251K} Kippenhahn, R., \&
Weigert, A.\ 1967, \zap, 65, 251

\bibitem[Landau
\& Lifshitz(1975)]{1975ctf..book.....L} Landau, L.~D., \&
Lifshitz, E.~M.\ 1975, Course of theoretical physics - Pergamon
International Library of Science, Technology, Engineering and
Social Studies, Oxford: Pergamon Press, 1975, 4th rev.engl.ed.,

\bibitem[Paczy{\'n}ski(1971)]{1971ARA&A...9..183P} Paczy{\'n}ski, B.\ 1971, \araa, 9, 183

\bibitem[Podsiadlowski et al.(2002)]{2002ApJ...565.1107P} Podsiadlowski,
P., Rappaport, S., \& Pfahl, E.~D.\ 2002, \apj, 565, 1107

\bibitem[Ritter(1988)]{1988A&A...202...93R} Ritter, H.\ 1988, \aap, 202, 93

\bibitem[Ritter(2008)]{2008NewAR..51..869R} Ritter, H.\ 2008, \nar, 51, 869

\bibitem[Stevens et al.(1992)]{1992MNRAS.254P..19S} Stevens, I.~R., Rees,
M.~J., \& Podsiadlowski, P.\ 1992, \mnras, 254, 19P

\bibitem[van Haaften et al.(2012)]{2012A&A...541A..22V} van Haaften, L.~M., Nelemans, G.,
Voss, R., \& Jonker, P.~G.\ 2012, \aap, 541, A22

\bibitem[Verbunt
\& Zwaan(1981)]{1981A&A...100L...7V} Verbunt, F., \& Zwaan, C.\ 1981, \aap, 100, L7

\end{thebibliography}
\end{document}